\font\subtit=cmr12
\font\name=cmr8

\def\plb#1#2#3#4{#1, {\it Phys. Lett.} {\bf {#2}}B (#3), #4}
\def\npb#1#2#3#4{#1, {\it Nucl. Phys.} {\bf B{#2}} (#3), #4}

\def\jmp#1#2#3#4{#1, {\it Jour. Math. Phys.} {\bf {#2}} (#3), #4}
\def\hpa#1#2#3#4{#1, {\it Helv. Phys. Acta} {\bf {#2}} (#3), #4}

\input harvmac
\def\UTN#1#2#3
{\TITLE{UTF-#1-\number\yearltd}
{#2}{#3}}
\def\TITLE#1#2#3{\nopagenumbers\abstractfont\hsize=\hstitle\rightline{#1}
\vskip 1in
\centerline{\subtit #2}
\vskip 1pt
\centerline{\subtit #3}\abstractfont\vskip .5in\pageno=0}%
\UTN{358}
{TOPOLOGICALLY NONTRIVIAL SECTORS OF MAXWELL}{FIELD THEORY ON RIEMANN
SURFACES}
\centerline{F{\name RANCO} F{\name ERRARI}}\smallskip
\centerline{\it Dipartimento di Fisica, Universit\'a di Trento, 38050 Povo,
Italy}
\centerline{\it and INFN, Gruppo Collegato di Trento}
\vskip 4cm
\centerline{ABSTRACT}
\vskip 1cm
{\narrower In this paper the Maxwell field theory is considered on a closed
and orientable Riemann surface of genus $h>1$. The solutions
of the Maxwell equations
corresponding to nontrivial values of the first Chern class are
explicitly constructed for any metric in terms of the prime form.}
\Date{August 1996}
%\draft
Two dimensional gauge field theories on manifolds have recently been
considered by various authors \ref\atiya{M. Atiyah and R. Bott,
{\it Phil. Trans. R. Soc. Lond.}
{\bf A 308} (1982), 523.}\nref\witten{
E. Witten, 
{\it Comm. Math. Phys.}
{\bf 141}
(1991),153;
{\it Jour. of Geom. and Phys.}
 {\bf 9}
(1992), 3781.}\nref\gross{D. J. Gross
{\it Nucl. Phys.}
{\bf B400} 
(1993), 161; D. J. Gross and Washington Taylor IV,
{\it Nucl. Phys.}
{\bf B400}
(1993), 181;
{\it ibid.}
{\bf B403}
(1993),
395}\nref\thompson{
%\jmp{
%M. Blau and G. Thompson}{36}{1995}{2192};
%M. Blau and G. Thompson,
%{\it Int. Jour. Mod. Phys.}
%{\bf A7}
%(1992), 3781; {\it Lectures on $2-d$ Gauge Theories}, lectures presented
%at the 1993 Trieste Summer School in High Energy Physics and Cosmology,
%Preprint IC/93/356,
%(hep-th/9310144);
G. Thompson, {\it 1992
Trieste Lectures on Topological Gauge Theory and Yang--Mills Theory},
Preprint IC/93/112 (hep-th/9305120).
%; D. Birmingham, M. Blau,
%M. Rakowski and G. Thompson, {\it Phys. Rep.} {\bf 209} (1991),
%129.
}\nref\bassetto{A. Bassetto, L. Griguolo and
G. Nardelli, {\it Yang--Mills Theories on the Space--Time
$S_1\times{\rm\bf R}$: Equal Time Quantization in Light Cone Gauge and
Wilson Loops}, Preprint DFPD-95-TH-07 (hep-th/9506095); \npb{A. Bassetto and
L. Griguolo}{439}{1995}{327}.}\nref\rajeev{\jmp{K. S. Gupta,
R. J. Henderson. S. G. Rajeev and O. T. Torgut}
{35} {1994}{3845}; S. G. Rajeev, {\it Phys. Lett.} {\bf 212B} (1988),
203.}\nref\wipf{
\hpa{I. Sachs and A. Wipf}{65}{1992}{653}; {\it Phys. Lett} {\bf 326B}
(1994), 105.}\nref\abdalla{
E. Abdalla, M.C.B. Abdalla and K. D. Rothe, {\it Nonperturbative 
Methods in Two Dimensional Quantum Field Theory}, Singapore 1991, World 
Scientific.}\nref\hitchin{
N. J. Hitchin, {\it Gauge Theories on Riemann Surfaces},
in Lectures on Riemann Surfaces, M. Cornalba \& al. (eds), ICTP Trieste Italy,
9 Nov.-18 Dec. 1987, World Scientific editions; {\it Topology}, {\bf 31}
(3) (1992), 449; {\it Proc. London 
Math. Soc.} {\bf 55}, 59 (1987).}\nref\fine{D. S. Fine,
{\it Comm. Math. Phys.}
{\bf 134}
(1990), 273; {\bf 140} (1991), 321;
{\it Quantum Yang--Mills on a Riemann Surface II:
Topological Sectors and Measures on Moduli Space},
(hep-th 9504103).}\nref\sengupta{A. Sengupta,
{\it Comm. Math.
Phys.} {\bf 147} (1992), 191; {\it Ann. Phys.} {\bf 221} (1993),
17.}\nref\joos{H. Joos, {\it Nucl. phys.} {\bf B17}
(Proc. Suppl.) (1990), 704; {\it Helv. Phys. Acta} {\bf 63} (1990),
670.}--\ref\jayewardena{\hpa{C. Jayewardena}{61}{1988}{636}.}.
In this paper the Maxwell field theory is treated on a closed and
orientable Riemann surface of genus $h>1$. The explicit construction of the
propagators and of the flat connections 
has been already discussed for the more general
Yang--Mills field theories in ref.
\ref\ffhpa{\hpa{F. Ferrari}{67}{1994}{702}.}.
However, also the abelian case is interesting because of the presence of
topologically nontrivial sectors.
For instance, the topologically nontrivial solutions of the
Maxwell equations have applications both
in string theory \ref\ketlec{\plb{S. V.
Ketov and O. Lechtenfeld}{353}{1995}{463}.} and in the Fractional Quantum
Hall Effect \ref\ienli{\npb{R. Iengo and Dingping Li}{413}{1994}{735}.}.
On Riemann surfaces equipped with arbitrary metrics
a rigorous and explicit
derivation of the gauge connections labelled
by nonzero values of the first
Chern class \ref\geometry{T. Eguchi, P. B. Gilkey and A. J. Hanson,
{\it Phys. Rep.} {\bf 66} (1980), 213.} is
still missing in the physical literature.
This derivation is
the main result of this letter.
To this purpose, we first
compute the general solutions of the Maxwell field theory
corresponding to a total magnetic flux $\Phi$ in terms of the prime form
\ref\fay{J. D. Fay, {\it Lect. Notes in Math.} {\bf 352}, Springer
Verlag, 1973.} and of the $2h$ holomorphic differentials. For general values
of $\Phi$, these solutions are periodic around the homology cycles of the
Riemann surface. However, we show here that the periodicity amounts
to a well defined gauge transformation is $\Phi$ satisfies the Dirac
quantization condition. With our formulas it is  possible for example
to explicitly verify the fact that the wave functions of the Landau levels
in the Quantum Hall effect are independent of the metric as indirectly
proved in \ienli.

Let us start from the action of the Maxwell field theory which is given
by:
\eqn\maxwell{S_{Maxwell}=
\int_Mdx^1\wedge dx^2{\sqrt{|g(x)|}\over 4}F_{\mu\nu}
F^{\mu\nu}}
where $F_{\mu\nu}=\partial_\mu A_\nu-\partial_\nu A_\mu$ and
$M$ is a general closed and orientable Riemann surface of genus $h>1$
parametrized by the real coordinates $x^\mu$, $\mu=1,2$. From now on we will
put $dx^1\wedge dx^2\equiv d^2x$ for simplicity.
On $M$ we define an Euclidean metric $g_{\mu\nu}(x)$ with determinant
$g(x)={\rm det}g_{\mu\nu}$ and
a canonical basis of homology cycles $a_i$ and $b_i$, with $i=1,\ldots,h$.
Let us now choose local complex coordinates $z=x^1+ix^2$ and $\bar z=x^1-ix^2$.
In these variables the determinant of the metric $g_{\alpha\beta}$,
$\alpha,\beta=z,\bar z$ is given by $g(z,\bar z)={1\over 4}g(x)$.
Explicitly, $g(z,\bar z)=g_{zz}g_{\bar z\bar z}-g_{z\bar z}g_{\bar z z}$.
The $h$ holomorphic
differentials admitted on $M$ will be denoted with the symbols $\omega_i(z)dz$,
$i=1,\ldots,h$. We assume that they are normalized in such a way that
$\oint_{a_i}dz\omega_j(z)=\delta_{ij}$. As a consequence, the matrix
$$\Omega_{ij}\equiv\oint_{b_i}\omega_j(z)dz$$
is the so--called period matrix.
For future
purposes, it will also be necessary to recall
the basics of the Hodge decomposition
on Riemann surfaces without boundary. Any vector field
$(A_z,A_{\bar z})$ can be decomposed
according to the Hodge decomposition theorem in coexact, exact and harmonic
components as follows:
\eqn\hodge{A_z=\partial_z\phi+\partial_z\rho+A^{\rm har}_z\qquad\qquad
A_{\bar z}=-\partial_{\bar z}\phi+
\partial_{\bar z}\rho+A^{\rm har}_{\bar z}}
The coexact and exact components are expressed using the two scalar
fields $\phi$ and $\rho$ respectively.
Here $\phi$ is purely complex, while $\rho$ is real.
$A^{\rm har}_z$ and $A^{\rm har}_{\bar z}$
take into account of the presence of the holomorphic differentials.
The decomposition \hodge\ is not invertible unless
\eqn\consistency{\int_Md^2z\sqrt{|g(z\bar z)|}\phi(z,\bar z)=\int_Md^2z
\sqrt{|g(z\bar z)|}\rho(z,\bar z)=0}
Here $d^2z\sqrt{g}\equiv i\sqrt{g}d\bar z\wedge dz$ is the volume form on $M$
in complex coordinates.
Let us now
compute the equations of motion coming from the action \maxwell. Noting
that in two
dimensions the only non--vanishing components of the field strength
$F_{\mu\nu}$ are
\eqn\fzzb{F_{z\bar z}=-F_{\bar z z}=\partial_zA_{\bar z}-
\partial_{\bar z}A_z}
an easy computation shows that
the gauge fields satisfy the following equations
of motion:
\eqn\eomone{\partial_{\bar z}\left[{1\over \sqrt{|g(z,\bar z)|}}
\left(\partial_{\bar z}
A_z-\partial_zA_{\bar z}\right)\right]=0}
\eqn\eomtwo{\partial_z\left[{1\over \sqrt{|g(z,\bar z)|}}
\left(\partial_zA_{\bar z}-
\partial_{\bar z}A_z\right)\right]=0}
The solutions of the above equations
are of the form \thompson, \ienli:
\eqn\solgen{F_{z\bar z}dz\wedge d\bar z={2\pi i\Phi
\sqrt{|g(z,\bar z)|}\over A}d\bar z\wedge dz}
where $A=\int_Md^2x\sqrt{|g(x)|}$ is the area
of the surface and $\Phi$ is a real constant representing the total magnetic
flux.
To show this is a simple exercise, but it will be useful to fix the notation.
Let us perform the change
of variables $(z,\bar z)\rightarrow(x^1,x^2)$ in the integral
$I=\int_M d^2z F_{z\bar z}$. Since $d^2z=-2id^2x$ and $F_{12}=-2iF_{z\bar z}$,
where $F_{12}=\partial_1A_2-\partial_2A_1$,
one finds that $I=\int_M d^2xF_{12}$. Thus $I$ is proportional to the first
Chern class defined by:
\eqn\realcc{k={1\over 2\pi}\int_Md^2xF_{12}\qquad\qquad\qquad k=0,\pm 1,\ldots}
On the other side, from eq. \solgen, we have that $I=2\pi i\Phi\int_M d^2z
\sqrt{|g(z,\bar z)|}{d\bar z\wedge dz\over A}$. Remembering that
$2\sqrt{g(z,\bar z)}=\sqrt{g(x^1,x^2)}$, we find the relation
$d^2z\sqrt{g(z,\bar z)}=d^2x\sqrt{g(x^1,x^2)}$.
As a consequence $I=2\pi\Phi$ and thus
we obtain the desired relation giving $\Phi$ as the total magnetic flux
in or outgoing from the Riemann surface:
\eqn\flux{2\pi\Phi=\int_M d^2xF_{12}}
Because of the Dirac quantization of the flux, the
topologically non-trivial solutions 
of the equations of motion \eomone--\eomtwo\
are characterized by the following integer values of $\Phi$:
\eqn\alphadef{\Phi= k\qquad\qquad\qquad k=0,\pm 1\pm 2,\ldots}
Eq. \solgen\ is equivalent to the following set of equations:
\eqn\emota{\partial_zA_{\bar z}={i\pi C_1\sqrt{|g(z,\bar z)|}\over A}
\qquad\qquad
\partial_{\bar z}A_z={i\pi C_2\sqrt{|g(z,\bar z)|}\over A}}
provided the fields $A_z$ and $A_{\bar z}$ are coexact forms.
For consistency with eq. \flux, the two constants
$C_1$ and $C_2$ must obey the following relation: $C_1-C_2=2\Phi$.
To solve eq. \emota\ explicitly in the gauge fields, we introduce the
following Green function \ref\vv{\npb{E. Verlinde and H. Verlinde}{288}{1987}
{357}.}:
\eqn\kzw{K(z,w)={\rm log}|E(z,w)|^2-2\pi\sum\limits_{i,j=1}^h
\left[{\rm
Im} \int_w^z\omega_j(s)ds\right]
\left({\rm Im}\enskip
\Omega\right)_{ij}^{-1}
\left[{\rm
Im} \int_{w}^z\omega_j(s)ds\right]}
where Im$\enskip a={a-\bar a\over 2i}$ denotes the imaginary part of a complex
number $a$. $E(z,w)$ is the prime form
\fay\ and $\left|{\rm Im}\enskip
\Omega\right|_{ij}^{-1}$ is the inverse of the matrix $
{\left(\Omega-\bar\Omega\right)_{ij}\over 2i}$.
$K(z,w)$ satisfies the equation:
\eqn\kzweq{\partial_z\partial_{\bar z} K(z,w)=
\pi\delta^{(2)}_{z\bar z}(z,w)-\pi\sum\limits_{i,j=1}^h
\omega_i(z)\left({\rm Im}\enskip
\Omega\right)_{ij}^{-1}\bar\omega_j(\bar z)}
We will show now that a particular solution of eqs. \emota, uniquely defined
modulo holomorphic connections, is:
\eqn\soljeden{A_z^{\rm I}={iC_1\over A}\int_M d^2w\partial_zK(z,w)
\sqrt{|g(w,\bar
w)|}+i\pi C_1\sum\limits_{i,j=1}^h \omega_i(z)\left( {\rm Im}\enskip
\Omega\right)^{-1}_{ij} \int_{\bar z_0}^{\bar z}\bar\omega_j(\bar
w)d\bar w}
\eqn\soldwa{A_{\bar z}^{\rm I}={iC_2\over A}\int_M d^2w\partial_{\bar z}
K(z,w)\sqrt{|g(w,\bar
w)|}+i\pi C_2\sum\limits_{i,j=1}^h \bar\omega_i(\bar z)\left({\rm Im}\enskip
\Omega\right)^{-1}_{ij} \int_{z_0}^z\omega_j(w)dw}
First of all, it is easy to see with the help of eq. \kzweq\ that the
above gauge
fields  really satisfy eqs. \emota.
Moreover,
since the field strength \fzzb\ is completely antisymmetric in its indices, it
takes no contributions from exact components. Indeed, the above vector fields
$A_{z}^{\rm I}$ and $A_{\bar z}^{\rm I}$ are purely coexact 1--forms.
To verify this, we note
that $A^I_z$ and $A_{\bar z}^I$ can be obtained, apart from a constant
factor, after derivation of the purely complex function
\eqn\fizzb{\phi^I(z,\bar z)={i\over A}\int_Md^2w\sqrt{|g(w,\bar w)|}K(z,w)+
i\pi\sum\limits_{i,j=1}^h\int_{z_0}^z\omega_i(w)dw\left({\rm Im}\enskip\Omega
\right)^{-1}_{ij}\int_{\bar z_0}^{\bar z}\bar \omega_j(\bar w)d\bar w}
Exploiting the symmetry properties of the period matrix
$\Omega$ in its indices, we have in fact:
\eqn\auxone{A_z^I=C_1\partial_z\phi^I\qquad\qquad\qquad
A_{\bar z}^I=C_2\partial_{\bar z}\phi^I}
Comparing with eq. \hodge, it is easy to see that the gauge fields
$A^I_\alpha$ of eqs. \soljeden--\soldwa\ are coexact only if
$C_1=-C_2$.
This condition, together with eq. \realcc, fixes the constants $C_1$
and $C_2$ uniquely:
\eqn\conectwo{C_1=-C_2=\Phi}
Finally we have to check that the particular solutions $A_z^I$ and
$A_{\bar z}^I$ given in \soljeden--\soldwa\ remain single-valued,
apart from a well defined gauge transformation, when transported around any
nontrivial
homology
cycle $\gamma$. Indeed, from their definitions \soljeden--\soldwa, it turns
out that the gauge fields $A_z^I$ and $A_{\bar z}^I$ are periodic around
$\gamma$:
\eqn\transfone{A_z^{\prime
I}=A_z^I+iC_1\pi\sum\limits_{i,j=1}^h\omega_i(z) \left({\rm
Im}\enskip\Omega\right)_{ij}\oint_\gamma\bar\omega(\bar w)d\bar w}
\eqn\transftwo{A_{\bar z}^{\prime
I}=A_{\bar z}^I+iC_2\pi\sum\limits_{i,j=1}^h\bar\omega_i(\bar z) \left({\rm
Im}\enskip\Omega\right)_{ij}\oint_\gamma\omega(w)dw}
These transformations amount however to a $U(1)$ gauge transformation
\eqn\gaugetr{A_\alpha^{\prime I}=A_\alpha^I+iU^{-1}(z,\bar
z)\partial_\alpha U(z,\bar z)\qquad\qquad \alpha=z,\bar z}
where the element of the gauge group is provided by:
$$U(z,\bar z)=U(z_0,\bar z_0){\rm exp}\left[\pi
C_1\sum\limits_{i,j=1}^h\left(\int_{z_0}^z\omega_i(s)ds\left({\rm
Im}\enskip\Omega\right)_{ij} \oint_{\gamma}\bar\omega_j(\bar w)d\bar
w-\right.\right.$$
\eqn\elofgroup{\left.\left.
\int_{\bar z_0}^{\bar z}\bar\omega_i(\bar s)d\bar s\left({\rm
Im}\Omega\right)_{ij}\oint_\gamma\omega_j(w)dw\right)\right]}
First of all,
substituting the above defined group element in eq. \gaugetr, one obtains
the transformations \transfone--\transftwo\ as expected. 
Moreover, it is easy to check that the exponent appearing in the right
hand side of eq. \elofgroup\ is purely imaginary. As a
consequence it turns out that
$\overline{U(z,\bar z)}=U^{-1}(z,\bar z)$.
However, this is still not enough to conlcude that $U(z,\bar z)$ is really
an element of $U(1)$, because one should also verify that $U(z,\bar
z)$ is not multivalued when transported around a nontrivial homology
cycle. To this purpose, let us first consider the case in which
$\gamma=b_l$, with $l=1,\ldots,h$.
After a slight reshuffling of indices, we have from
eq. \elofgroup:
$$\left.U(z,\bar z)\right|_{\gamma=b_l}=\left.U(z_0,\bar
z_0)\right|_{\gamma=b_l}{\rm exp}\left[2\pi i
C_1\sum\limits_{i,j=1}^h\left(\bar\Omega_{li}\left(\Omega-\bar\Omega
\right)^{-1}_{ij}\int_{z_0}^z\omega_i(s)ds-\right.\right.$$$$\left.\left.
\Omega_{li}\left(\Omega-\bar\Omega
\right)^{-1}_{ij}\int_{\bar z_0}^{\bar z}\bar\omega_i(\bar s)d\bar
s\right)\right]$$
At this point we recall that the real
harmonic differentials $\alpha_l$, defined in such a way that
\eqn\hardiffone{\oint_{a_k}\alpha_l=\delta_{kl}\qquad\qquad\qquad
\oint_{b_k}\alpha_l=0}
have the following expression \ref\farkra{H. M. Farkas and I. Kra, {\it
Riemann Surfaces}, Springer Verlag, Berlin, Hedelberg, New York 1980.}:
$$\alpha_l=\sum\limits_{i,j=1}^h\left(-\bar\Omega_{li}(\Omega-
\bar\Omega)^{-1}_{ij}\omega(z)dz+c.c.\right)$$
where $c.c.$ stands for complex conjugate.
Therefore, setting
$z_0=(x_0^1,x_0^2)$ and $z=(x^1,x^2)$, $U(z,\bar z)$ becomes in real
coordinates:
$$\left.U(x)\right|_{\gamma=b_l}=\left.U(x_0)\right|_{\gamma=b_l}{\rm
exp} \left[-2\pi i C_1\int_{(x_0^1,x_0^2)}^{(x^1,x^2)}\alpha_l\right]$$
Here we have used the relation $\int_\sigma dx^\mu A_\mu=
\int_\sigma(dzA_z+d\bar zA_{\bar
z})$, valid for any vector field $A_\mu$ and
for any path $\sigma$ on $M$.
Now we move $U(x)$ along a closed cycle $b_k$. Let us denote the
result of this operation with $U'(x)$. Since the exponent in
the right hand
side of eq. \elofgroup\ is purely imaginary, $U'(x)$ should differ from
$U(x)$ at most by a complex phase. However, exploiting
eqs. \hardiffone, it is easy to see that:
$$U'(x)=U(x)$$
Analogously, transporting $U(x)$ around a cycle $a_k$, we have:
$$U'(x)=U(x)e^{-2\pi iC_1\delta_{kl}}$$
Remembering the condition \alphadef\ on the magnetic flux and eq. \conectwo,
it is clear that $e^{-2\pi
iC_1\delta_{kl}}=1$,
proving that $U(z,\bar z)$ is singlevalued around the
$b-$cycles.

To conclude our proof, we consider the case in which
$\gamma=a_l$ in eq. \elofgroup.
Performing similar calculations as before, one finds:
$$\left.U(x)\right|_{\gamma=a_l}=\left.U(x_0)\right|_{\gamma=a_l}{\rm
exp} \left[2\pi i C_1\int_{(x_0^1,x_0^2)}^{(x^1,x^2)}\beta_l\right]$$
where the symbols $\beta_l$, $l=1,\ldots,h$,
denote now the real  harmonic differentials
$$\beta_l=\sum\limits_{i=1}^h\left[(\Omega-\bar\Omega)^{-1}_{li}\omega_i(z)dz-
(\Omega-\bar\Omega)^{-1}_{li}\bar\omega_i(\bar z)d\bar z\right]$$
They are normalized along the homology cycles as follows \farkra:
$$\oint_{a_k}\beta_l=0\qquad\qquad\qquad\oint_{b_k}\beta_l=\delta_{kl}$$
Exploiting the above equations, a straightforward calculation
shows that  $U(x)$ is
singlevalued also when transported across an $a-$cycle. This completes our
proof that $U(z,\bar z)$ is a well defined element of the group
$U(1)$. As a consequence, the claim has been verified, that the gauge fields
given in \soljeden--\soldwa\ are particular solutions of the Maxwell
equations \eomone--\eomtwo\ on any closed and orientable Riemann surface
$M$ equipped with an arbitrary metric $g_{\mu\nu}$.
These solutions correspond to topologically non-trivial configurations
with nonvanishing first Chern class.
If the total flux $\Phi$ is not an integer, the periodicity of $A_\alpha^I$
around the
homology cycles, computed in eqs. \transfone--\transftwo, cannot be reabsorbed
by a gauge transformation, since $U(z,\bar z)$ starts to be multivalued and
thus it is no longer a valid group element on $M$.
All the other solutions corresponding to nonvanishing
values of the first Chern class, can be obtained from eqs. \soljeden--\soldwa\
adding linear combinations of the harmonic differentials and exact forms.

Concluding, we have computed here the nontrivial solutions of the
Maxwell equations. The Dirac quantization of the total flux $\Phi$
naturally arises from the requirement that the periodicity of these solutions
around the homology cycles can be reabsorbed by a well defined
gauge transformation. As
briefly mentioned in the Introduction, there are some applications
of our results both in the quantum Hall effect and in the theory of
superstrings with $N=2$. These topics lay however beyond the scope of
this short letter and will be treated elsewhere.

%
% Let us also briefly mention that the topologically
%nontrivial sectors of the Maxwell field theory have also been derived
%on the $Z_n$ symmetric algebraic curves using a totally different
%formalism confirming the results presented here
%\ref\ffnew{F. Ferrari,
%{\it Topologically NonTrivial Sectors of the Maxwell Field
%Theory on Algebraic Curves}, Preprint UTF 379/96.}.
\listrefs
\bye